\documentstyle[multicol,preprint,aps,epsf]{revtex}
\begin{document}
\title{Coarsening and Persistence in the Voter Model}
\author{E.~Ben-Naim$^1$, L.~Frachebourg$^2$, and P.~L.~Krapivsky$^3$}
\address{$^1$The James Franck Institute,
The University of Chicago, Chicago, IL 60637}
\address{$^2$Center for Polymer Studies and Department of Physics,
Boston University, Boston, MA 02215}
\address{$^3$Courant Institute of Mathematical Sciences,
New York University, New York, NY 10012}
\maketitle
\begin{abstract}
We investigate coarsening and persistence in the voter model by
introducing the quantity $P_n(t)$, defined as the fraction of voters
who changed their opinion $n$ times up to time $t$. We show that
$P_n(t)$ exhibits scaling behavior that strongly depends on the
dimension as well as on the initial opinion concentrations.  Exact
results are obtained for the average number of opinion changes,
$\langle n\rangle$, and the autocorrelation function, $A(t)\equiv \sum
(-1)^nP_n\sim t^{-d/2}$ in arbitrary dimension $d$.  These exact
results are complemented by a mean-field theory, heuristic arguments
and numerical simulations.  For dimensions $d>2$, the system does not
coarsen, and the opinion changes follow a nearly Poissonian
distribution, in agreement with mean-field theory. For dimensions
$d\leq 2$, the distribution is given by a different scaling form,
which is characterized by nontrivial scaling exponents. For unequal
opinion concentrations, an unusual situation occurs where different
scaling functions correspond to the majority and the minority, as well
as for even and odd $n$.
\end{abstract}


\section{Introduction}

The theory of phase separation, or domain coarsening, has undergone a
significant development in the last three decades \cite{bray}. The
most important finding is that well-defined ordered domains arise and
grow with time in such a way that the coarsening process exhibits
scaling.  In other words, at the late stages of the evolution the
system is characterized by a single length scale $L(t)$ that gives a
typical linear size of the domains. It is well established, at least
for systems with a scalar order parameter, that $L(t)\sim t^n$ with
$n=1/2$ for {\it non-conserved} dynamics and $n=1/3$ for {\it
conserved} dynamics. For the Ising spin model, Glauber spin-flip
dynamics exemplifies the former, while Kawasaki spin-exchange dynamics
exemplifies the later.

Several important correlation functions exist.  One such function, the
autocorrelation or, equivalently, the two-time equal-space correlation
function, $A(t)$, is defined by $A(t)=\langle \phi({\bf r},0)\phi({\bf
r},t)\rangle$, where $\phi({\bf r},t)$ is the order parameter.  Then,
scaling implies: $A(t)\sim L^{-\lambda}(t)$, with an exponent
$\lambda$ \cite{fisher}.  The general two-point correlation function,
$g(r,t)=\langle \phi({\bf 0},0)\phi({\bf r},t)\rangle$, can be
expressed through $\lambda$, namely $g(r,t)=L^{-\lambda}G(r/L)$.
Exact results for $\lambda$ are known in few cases
\cite{glauber,cz,jan,derrida}, while the bounds $d/2\leq \lambda\leq
d$ with $d$ the spatial dimension were proposed by Fisher and Huse
\cite{fisher}.  For the $O(m)$ vector model in the $m\to \infty$
limit, $\lambda=d/2$ \cite{cz}. In this study, the value $\lambda=d$
is obtained for the voter model, defined below.  This result indicates
that both the upper and the lower bounds can be realized.

It should be noted that in most coarsening processes the dynamics does
not exhibit a qualitative dependence on the temperature $T$ as long as
$T<T_c$ \cite{bray,bray1}. At the critical temperature $T=T_c$, the
dynamics is generally different, and ordered domains usually do not
occur. However, the correlation length $\xi(t)$ exists and grows with
time as $\xi(t)\sim t^{1/z}$, where $z$ is the dynamical exponent. The
correlation length $\xi(t)$ should be considered as the analog of the
domain size $L(t)$ \cite{jan,huse}, and the the exponent $\lambda$ is
replaced by $\lambda_c$ defined by $A_c(t)\sim\xi^{-\lambda_c}$. In
the voter model, temperature is absent but since the dynamics is
noiseless, the voter model dynamics is zero temperature in nature.
However, the ``critical'' temperature is also zero.  If one introduces
noise by allowing environment-independent opinion changes, the system
does not coarsen (see, e.g., \cite{kr}). Thus, we will actually
establish $\lambda_c=d$ for the voter model.  A general discussion of
the conditions under which the equality $\lambda=\lambda_c=d$ holds is
given by Majumdar and Huse \cite{mhuse}.

In this study, we introduce a family of quantities which provides
insight into the ``history'' the the coarsening process. We denote by
$P_n(t)$ the fraction of voters who changed their opinion exactly $n$
times during the time interval $(0,t)$.  The first of these
quantities, $P_0(t)$, is equal to the fraction of persistent voters,
{\it i.e.} voters who did not change their opinion up to time
$t$. This quantity has been introduced independently for two
equivalent one-dimensional models, the Glauber-Ising model \cite{dbg},
and the single-species annihilation process \cite{kbr}.  Furthermore,
the corresponding generalizations to arbitrary dimensions were
discussed in \cite{ising} and \cite{annih}, respectively.  Derrida
{\it et al.} \cite{der} established the exact asymptotic decay of this
quantity, $P_0(t)\sim t^{-3/8}$, as suggested earlier by numerical
simulations.  Another exact result \cite{bdg} establishes $P_0(t)\sim
L^{-\beta}$, with $\beta=0.175075\ldots$, for the 1D time-dependent
Ginsburg-Landau equation at zero temperature.

For the voter model, several quantities such as the one-time and the
two-time correlations are exactly solvable in arbitrary
dimensions \cite{kr,lp}.
These correlation functions allow an exact calculation of the average
number of opinion changes $\langle n \rangle$ and other interesting
quantities.  Hence, the voter model is a natural starting point for
investigation of $P_n(t)$. Although we do not obtain the full
distribution, most of its features are illuminated by combining the
above exact results with heuristic arguments and with the mean-field
solution.  Generally, $P_n(t)$ exhibits a scaling behavior. For $d>2$,
the scaling function is Poissonian and is peaked at $n=\langle n
\rangle$, while for $d=1$ the distribution is maximal near the origin.
Additionally, for unequal opinion concentrations different scaling
functions for even and odd opinion changes are found.  Using random
walks techniques, we obtain the even and odd scaling functions in the
limit of an infinitesimal minority concentration.

The rest of this paper is organized as follows. In Sec.~II, we first
solve for $P_n(t)$ on a complete graph. We then reexpress some exact
relationships for the voter model, in arbitrary dimension $d$, in
terms of the distribution $P_n(t)$. Combining with the mean-field
solution, these exact relationships allow us to guess the scaling form
of $P_n(t)$.  This guess suggests a usual scaling form in one
dimension, and a mean-field like sharply peaked distribution for
$d>1$. These predictions are then compared with numerical data in one,
two, and three dimensions. In Sec.~III, we describe exact solution of
the mean-field equations for the case of initially different
concentrations. Then we present exact results for the autocorrelation
function in arbitrary dimension $d$, and exact results for the
fraction of persistent voters $P_0(t)$ in one dimension. We proceed by
investigating the extreme case of infinitesimal minority opinion. In
this limit, the model is equivalent to a pair of annihilating random
walkers who are nearest neighbor at $t=0$. Simplifying further the
problem to the case of a random walker near the absorbing boundary we
derive a complete analytical solution. Finally, we perform numerical
simulations for the case of unequal concentrations and confront the
results to exact predictions.  Finally, we give a brief summary in
Sec.~IV.

\section{Equal Concentrations}

In this section we first define the voter model. We restrict attention
to the symmetric case, {\it i.e.}, equal opinion densities. We start
by analyzing the mean-field theory of the model, and then obtain
several exact results in arbitrary dimensions. We then present a
scaling ansatz and check it using numerical simulations.

\subsection{Mean-Field Theory}

We start by defining the voter model \cite{lig}.  Consider an
arbitrary lattice and assume that each site is occupied by a ``voter''
who may have one of two opinions, denoted by $+$ and $-$.  Each site
keeps its opinion during some time interval, distributed exponentially
with a characteristic time $\tau$, set to unity for convenience, and
then assumes the opinion of a randomly chosen neighboring site.  If a
site is surrounded by sites with the same opinion, it does not change
its opinion. Hence, such dynamics are zero-temperature in
nature. Noise can be introduced by allowing a voter to change its
opinion independently of its neighbors.  However, a voter system with
noisy dynamics does not coarsen, and we restrict ourselves to the
noiseless voter dynamics. These dynamics are so simple that they
naturally arise in a variety of situations, see
e.g. \cite{kr,lig}. An important link is with the Glauber-Ising model:
In one dimension, and {\it only} in 1D, the voter dynamics is
identical to the Glauber dynamics. This equivalence is not restricted
to zero temperature, 1D noisy voter dynamics is also identical to
Glauber dynamics at a positive temperature.

We now consider the voter model dynamics on a mean-field level, by
simply treating all sites as neighbors. Such a theory is {\it exact}
on a complete graph. Moreover, it is expected to hold in sufficiently
large spatial dimensions.  We first consider the symmetric case were
the opinions concentrations, $c_+$ and $c_-$,  are equal, and the
interesting case of unequal concentrations will be discussed later.
The fraction of sites which have changed their opinions $n$ times up
to time $t$, evolves according to
\begin{equation}
\label{mfe}
{dP_n\over dt}=P_{n-1}-P_n,
\end{equation}
with $P_{-1}\equiv 0$ to ensure $dP_0/dt=-P_0$. The distribution
clearly satisfies the normalization condition, $\sum_n P_n=1$, and one
can verify that Eq.~(\ref{mfe}) conserves this sum.  Solving
Eq.~(\ref{mfe}) subject to the initial condition $P_n(0)=\delta_{n0}$,
one finds that the opinion change distribution function is Poissonian
\begin{equation}
\label{mfs}
P_n(t)={t^n\over n!}\,e^{-t}.
\end{equation}
In particular, the fraction of persistent voters, {\it i.e.} voters
who did not change their opinion up to time $t$, decreases exponentially,
$P_0(t)=e^{-t}$.  The probability that a voter has its initial
opinion at time $t$ is thus $P_{\rm even}=\sum_n P_{2n}=(1+e^{-2t})/2$.
Asymptotically, this probability exponentially approaches the value
$1/2$, and therefore voters quickly ``forget'' their initial opinion.

The distribution is peaked around the average $\langle n \rangle=t$,
and the width of the distribution, $\sigma$, is given by
$\sigma^2=\langle n^2 \rangle-\langle n\rangle=t$.  In the limits,
$t\to \infty$, $n\to \infty$, and $(n-t)/\sqrt{t}$ finite, $P_n(t)$
approaches a scaling form
\begin{equation}
\label{mfscaling}
P_n(t)={1\over\sigma}
\Phi_{\infty}\left({n-\langle n\rangle\over\sigma}\right),
\end{equation}
where the scaling distribution is Gaussian
$\Phi_{\infty}(x)=(2\pi)^{-1/2} \exp(-x^2/2)$.  This
infinite-dimension scaling solution will be compared below to
simulation results in three dimensions. To summarize, the quantity
$P_n(t)$ incorporates many statistical
properties of the system such as the probability of maintaining the
original opinion, the probability of having the original opinion, and
the average number of opinion changes.

\subsection{Exact Results}

We now review several relevant known exact results for the voter model
in arbitrary dimension $d$ and reexpress them in terms of $P_n(t)$.
Both the one- and two-body equal-time correlation functions
\cite{kr,lp}, are exactly solvable on arbitrary lattice in arbitrary
dimension.  It proves useful to formulate the voter model
on the language of Ising spins, {\it i.e.},
a $+$ opinion is identified with  $+1$ spin and a $-$ opinion with $-1$
spin.  The state of the lattice is described by $S\equiv[S_{{\bf k}}]$,
the spin-flip rate $W_{{\bf k}}(S)\equiv W(S_{{\bf k}}\to -S_{{\bf k}})$ reads
\begin{equation}
      \label{rate}
W_{{\bf k}}(S)=1-{1\over z_d}
S_{{\bf k}}\sum_{{\bf e}_i}S_{{\bf k+e}_i},
\end{equation}
with $z_d$ the coordination number.  Here the sum in the right-hand
side runs over all $z_d$ nearest neighbors. It is convenient to
rescale the time variable, $t\to t z_d/4$.  The probability
distribution $P(S,t)$ satisfies the master equation
\begin{equation}
      \label{master}
{d\over dt}P(S,t)=\sum_{{\bf k}}\left[W_{{\bf k}}(S^{{\bf k}})
P(S^{{\bf k}},t)-W_{{\bf k}}(S)P(S,t)\right],
\end{equation}
where the state $S^{{\bf k}}$ differs from $S$ only at the site {\bf
k}.  One can then derive a set of differential equations for the spin
correlation functions $\langle S_{{\bf k}}\ldots S_{{\bf l}}\rangle
\equiv \sum_{S}S_{{\bf k}}\ldots S_{{\bf l}}P(S,t)$.  The single- and
two-body correlation functions satisfy discrete Laplace
equations \cite{kr},
\begin{eqnarray}
\label{1body}
2{d\over dt}\langle S_{{\bf k}}\rangle&=&-z_d\langle S_{\bf k}\rangle
+\sum_{{\bf e}_i}\langle S_{{\bf k+e}_i}\rangle,\\
2{d\over dt}\langle S_{\bf k}S_{\bf l}\rangle&=&
-2z_d\langle S_{\bf k}S_{\bf l}\rangle
+\sum_{{\bf e}_i}\langle S_{{\bf k+e}_i}S_{\bf l}\rangle
+\sum_{{\bf e}_i}\langle S_{\bf k}S_{{\bf l+e}_i}\rangle.\nonumber
\end{eqnarray}

On a simple (hyper)cubic lattice where $z_d=2d$, the general
solution for the average opinion is given by
\begin{equation}
      \label{1sol}
\langle S_{{\bf k}}\rangle=e^{-td}\sum_{\bf l}
\langle S_{\bf l}(0)\rangle\,I_{{\bf k}-{\bf l}}(t),
\end{equation}
where $I_{\bf k}(x)$ is the multi-index function, $I_{\bf
k}(x)=\prod_{1\leq j\leq d}I_{k_j}(x)$, and
$I_n$ is the modified Bessel function.  The autocorrelation
$A(t)=\langle S_{\bf 0}(0) S_{\bf 0}(t)\rangle$ is of particular
interest since it is related to the opinion change distribution via
the alternating sum $A(t)=\sum (-1)^nP_n(t)$. The autocorrelation is
found from Eq.~(\ref{1body}), $A(t)=e^{-td}\sum_{\bf l}\langle S_{\bf 0}(0)
S_{\bf l}(0)\rangle\,I_{{\bf l}}(t)$.  In the simplest case of
completely uncorrelated initial opinions, with equal densities of the
opposite opinions, $\langle S_{\bf 0}(0) S_{\bf
l}(0)\rangle=\delta_{{\bf l0}}$, one finds
\begin{equation}
\label{auto}
A(t)=\sum_n (-1)^nP_n(t)=\left[e^{-t}I_0(t)\right]^d,
\end{equation}
and thus asymptotically, $A(t)\simeq (2\pi t)^{-d/2}$.
The diffusive
nature of the problem (see e.g. Eq.~(\ref{1body})) suggests that the
correlation length is given by the diffusion scale,
$\xi(t)\sim\sqrt{t}$. Therefore, the autocorrelation function scales
as $\xi^{-d}$ for arbitrary $d$, thus implying that the exponent
$\lambda_c$ is well defined in all dimensions, and equal to
$\lambda_c=d$ as claimed previously.

The average number of opinion changes, $\langle n\rangle=\sum_n n P_n$,
is simply related to the concentration of ``active bonds''
(neighbors with different opinions)
$c_{+-}\equiv (1-\langle S_{\bf l}S_{{\bf l}+{\bf e}}\rangle)/2$:
$d\langle n\rangle/dt=c_{+-}$.
Evaluation of the active bonds density gives the
following leading asymptotic behavior \cite{lp}
\begin{equation}
\label{inter}
c_{+-}(t)\sim\cases
{t^{-1+d/2}, & $d<2$, \cr
1/\ln t,     &  $d=2$, \cr
{\rm const}, & $d>2$. \cr}
\end{equation}
Thus, when $d\leq 2$, the density of active bonds vanishes for
sufficiently long-time, {\it i.e.}  coarsening takes place in low
dimensions.  In contrast, for $d>2$, single-opinion domains do not
arise. This is not very surprising since at the critical point
well-ordered domains should not necessarily form.
Following the above
discussion, the average number of opinion changes in the limit
$t\to\infty$ is obtained by integrating $c_{+-}$,
\begin{equation}
\label{sum2}
\langle n\rangle\sim\cases
{t^{d/2}, &  $d<2$, \cr
t/\ln t,  &  $d=2$, \cr
t,        &  $d>2$. \cr}
\end{equation}

The above results agree with the mean-field results when $d>2$.
We therefore expect that for $d>2$, the distribution function $P_n(t)$
approaches the Poissonian distribution of Eq.~(\ref{mfscaling}).
Similarly, the fraction of persistent voters $P_0(t)$ should decay
exponentially in time as well.
Interestingly, the exact result for
the autocorrelation function indicates a subtle failure of the
mean-field approach concerning the probability that a voter has
its initial opinion $P_{\rm even}(t)=\sum P_{2n}(t)=\left(1+A(t)\right)/2$.
{}From Eq.~(\ref{auto}), one finds that
asymptotically $P_{\rm even}(t)-1/2\sim t^{-d/2}$, while the mean-field
approach gives $P_{\rm even}(t)-1/2\sim e^{-2t}$. Hence voters have a
stronger than exponential memory, even for $d>2$.  Despite this
erroneous prediction, the mean-field theory is successful in predicting
most features of the opinion change distribution function for $d>2$.

\subsection{Scaling Arguments}

We were unable to find the exact $P_n(t)$ distribution for $d\le
2$, or higher moments such as $\langle n^2\rangle$. However, combining
the above results with scaling arguments proves useful. In
one dimension, the average number of changes scales as $\langle
n\rangle \sim \sqrt{t}$. We assume that this scale characterizes the
distribution, or in other words, $\sigma \sim \langle n\rangle\sim
\sqrt{t}$. Thus, we arrive at the scaling form
\begin{equation}
\label{1dscaling}
P_n(t)={1\over \sqrt{t}}\,\Phi_1\left({n\over \sqrt{t}}\right).
\end{equation}
In general we will use the notation $\Phi_d$ for the $d$-dimensional
scaling function. The nontrivial decay of the persistent voter
density, $P_0\sim t^{-3/8}$\cite{der}, implies nontrivial divergence of the
scaling form $\Phi_1(z)\sim z^{-1/4}$, in the small argument limit,
$z\to 0$. The tail of the distribution corresponds to a large number
of opinion changes by a specific voter and can be estimated by an
intuitive argument.  Such a voter must reside at the boundary between
two single opinion regimes, and must change its mind constantly. The
probability that such a voter changes its mind $t$ times (one time per
unit time), can be estimated by $P_t(t)=\exp(-{\rm const}\times
t)$. It is also natural to assume that the scaling function rapidly
decays for large $z$, $\Phi_1(z)\sim\exp(-{\rm const}\times z^\alpha)$.
Combining this form with Eq.~(\ref{1dscaling}) gives
$P_t(t)\sim\exp(-{\rm const}\times t^{\alpha/2})$ and consequently,
$\alpha=2$. To summarize, the limiting behavior of the one-dimensional
scaling function are
\begin{equation}
\label{1dlimits}
\Phi_1(z)\sim \cases
{z^{-1/4}&$z\ll 1;$\cr
\exp(-{\rm const}\times z^2)&$z\gg1$.\cr}
\end{equation}
This scaling behavior is different in nature than the scaling behavior
for dimensions $d>2$. While for $d\geq 2$, a well-defined peak in
the distribution function occurs near the average, the one dimensional
distribution is a decreasing function of $n$. Moreover, the Gaussian
function $\Phi_{\infty}(z)=(2\pi)^{-1/2}\exp(-z^2)$ is symmetric
around its average, while no such symmetry occurs for $d=1$, as the
distribution is peaked near $n=0$. Despite these differences, the tail
of the distribution of Eq.~(\ref{1dlimits}) agrees with the mean-field
distribution of Eq.~(\ref{mfscaling}). In fact, the above heuristic
argument for the large-$n$ behavior is valid in arbitrary dimensions.

\subsection{Simulation Results}

We implement the voter model using a simple Monte-Carlo simulation. A
simple (hyper)cubic lattice is chosen with a linear size $L$, and
periodic boundary conditions are imposed.  A simulation step consists
of choosing randomly an active bond ({\it i.e.}, a bond between
neighbors with different opinions) and changing the opinion of one of
the two voters. After each such step, time is incremented by the
inverse number of active bonds and the active bond list is
updated. This simulation procedure is efficient for spatial dimensions
$d\le2$ since the system coarsens and the number of active bonds
decreases as the simulation proceeds.  The results below correspond to
one realization on a lattice of linear dimension $L=10^7$, $10^3$, and
$2\times10^2$ in 1D, 2D, and 3D, respectively.

In one dimension, the numerical results confirm the scaling ansatz of
Eq.~(\ref{1dscaling}), as shown in Fig.~1. Interestingly, the maximum
of the distribution occurs at $n=1$, and the distribution decays
monotonously for $n>1$. The postulated limiting behaviors of the
scaling distribution are also confirmed. To test
the validity of the mean-field theory, we performed numerical
simulations in three dimensions. The resulting $P_n(t)$ distribution
agrees with the Poissonian distribution of Eq.~(\ref{mfs}), and
furthermore the fraction of persistent voters decays exponentially.
These results indicate that the above mentioned discrepancy regarding
the autocorrelation function is not crucial in understanding the
opinion change distribution function.

The marginal two-dimensional case is especially interesting.  While it
is expected that the distribution will be roughly Poissonian, some
deviation from the mean-field predictions are expected. We find
numerically that the distribution obeys the scaling form of
Eq.~(\ref{mfscaling}), and exhibits a well defined peak in the
vicinity of $\langle n\rangle\sim t/ \ln t$.  However, in 2D, the
system still coarsens, and the distribution exhibits some low-
dimensional features. The distribution is not a symmetric function of
the variable $n-\langle n\rangle$ (Fig.~2). Additionally, an intriguing
behavior for the fraction of persistent voters is found numerically
(Fig.~3),
\begin{equation}
  \label{p0}
P_0(t)\sim \exp(-{\rm const}\times\ln^2t).
\end{equation}
Thus the fraction of persistent voters decays faster than the 1D
power-law behavior, and slower than the mean-field exponential
behavior and
a naive logarithmic correction to the mean-field behavior does not hold.
%A naive argument which combines the mean-filed behavior
%$p_0\sim \exp(-\langle n\rangle)$ with $\langle n\rangle\sim t/\ln t$
%gives $\exp(-{\rm const}\times t/\ln t)$ rather than Eq.~(\ref{p0}).
The width of the distribution $\sigma$ is not given by a
logarithmic correction to the mean-field (where $\sigma\sim t^{1/2}$),
and our best fit gives $\sigma\sim t/\ln^{\alpha} t$ with $\alpha>1$.
The 2D case is difficult from a numerical point of view since logarithmic
corrections occur, and a large temporal range is required to
distinguish such slowly varying corrections from algebraic behavior
with small exponents \cite{lp}.

\section{Unequal concentrations}

Our previous exposition has assumed that the initial concentrations of
dissimilar species are equal, $c_+=c_-=1/2$. The case of unequal
concentrations, $c_+\neq c_-$, is interesting as well.  The reason is
that the voter model dynamics has a remarkable feature: Although {\it
locally} the opinion does change (the dynamics is non-conserved in
nature), {\it globally} both opinions are conserved.  On the language
of the Ising model it can be said that at zero temperature the
magnetization remains constant. This hidden integral leads to several
peculiarities which will be illuminated first on a mean-field level
and then for a special case in one dimension.

\subsection{Mean-Field Theory}

 To write the general mean-field theory, it is necessary to distinguish
between voters according to their initial opinions. Hence, we
introduce $P_n^+(t)$ ($P_n^-(t)$), the fraction of voters with the $+$
($-$) initial opinion that have changed their opinion $n$ times up to
time $t$.  If all sites are neighbors, these distributions evolve
according to
\begin{eqnarray}
\label{mf}
{dP_{2n}^+\over dt}&=&2\left(c_+P_{2n-1}^+ -c_-P_{2n}^+\right),\nonumber\\
{dP_{2n+1}^+\over dt}&=&2\left(c_-P_{2n}^+ - c_+P_{2n+1}^+\right),\nonumber\\
{dP_{2n}^-\over dt}&=&2\left(c_-P_{2n-1}^- -c_+P_{2n}^-\right), \nonumber\\
{dP_{2n+1}^-\over dt}&=&2\left(c_+P_{2n}^- - c_-P_{2n+1}^-\right),
\end{eqnarray}
with $P_{-1}^{\pm}\equiv 0$.
The initial conditions are $P_n^{\pm}(t=0)=c_{\pm} \delta_{n0}$.  These
rate equations reduce to Eq.~(\ref{mfe}) for the symmetric case
$c_+=c_-=1/2$.

It is again useful to consider the fraction of voters that have (do
not have) their initial opinion, $P_{\rm even}^{\pm}=\sum_n
P_{2n}^{\pm}$ ($P_{\rm odd}^{\pm}=\sum_n P_{2n+1}^{\pm}$).  Summation
of Eqs.~(\ref{mf}) gives
\begin{equation}
  \label{ppmeo}
{dP_{\rm even}^{\pm}\over dt}=-{dP_{\rm
odd}^{\pm}\over dt}= 2c_{\pm}^2-2P_{\rm even}^{\pm}.
\end{equation}
One can find that the global
opinion concentrations $c_{\pm}=P_{\rm even}^{\pm} +P_{\rm odd}^{\mp}$
are conserved and that  $P_{\rm odd}^+=P_{\rm odd}^-$.
Solving these last rate equations subject to the proper
initial conditions gives
\begin{eqnarray}
\label{neven}
P_{\rm even}^{\pm}&=&c_{\pm}\left(c_{\pm}+c_{\mp}e^{-2t}\right),
\nonumber\\ P_{\rm odd}^{\pm}&&=c_+c_-\left(1-e^{-2t}\right).
\end{eqnarray}
The autocorrelation function is then given by
\begin{eqnarray}
\label{automf}
A(t)&=&P_{\rm even}^++P_{\rm even}^--P_{\rm odd}^+-P_{\rm odd}^-\nonumber\\
&&=(c_+-c_-)^2+4c_+c_-\,e^{-2t}.
\end{eqnarray}
A voter quickly forgets its initial opinion even if statistically it
is more likely to have its initial opinion since  $c_+^2+c_-^2\geq 2c_+c_-$.

The fraction of persistent voters is found by solving
$dP_0^{\pm}/dt=-2c_{\mp}P_0^{\pm}$ and it is  found that
\begin{equation}
\label{p0mf}
P_0^{\pm}=c_{\pm}\,e^{-2c_{\mp}t}.
\end{equation}
Thus, the fraction of persistent voters decays exponentially as well. The
decay constant is simply given by the density of opposite opinion. This
result indicates that even in the case of a small concentration of one
opinion, the fraction of persistent majority voters decays
exponentially with time.

To solve Eqs.~(\ref{mf}) we introduce the generating functions
\begin{eqnarray}
      \label{generating}
F_{\rm even}^{\pm}(t,w)&=&\sum_{n=0}^\infty  P_{2n}^{\pm}(t)w^{2n}, \nonumber\\
F_{\rm odd}^{\pm}(t,w)&=&\sum_{n=0}^\infty  P_{2n+1}^{\pm}(t)w^{2n+1}.
\end{eqnarray}
This reduces the infinite set of rate equations to four equations
\begin{eqnarray}
      \label{mfg}
{dF_{\rm even}^{\pm}\over dt}&=&
2\left(c_{\pm}wF_{\rm odd}^{\pm}-c_{\mp}F_{\rm even}^{\pm}\right), \nonumber\\
{dF_{\rm odd}^{\pm}\over dt}&=&
2\left(c_{\mp}wF_{\rm even}^{\pm}-c_{\pm}F_{\rm odd}^{\pm}\right).
\end{eqnarray}
Expressing $F_{\rm odd}^{\pm}$ via $F_{\rm even}^{\pm}$, we reduce the
system of first-order differential equations (\ref{mfg}) to a pair of
second-order equations for $F_{\rm even}^+(t,w)$ and $F_{\rm
even}^-(t,w)$.  Solving these equations subject to the proper boundary
conditions yields
\begin{equation}
      \label{gen}
F_{\rm even}^{\pm}(t,w)=
c_{\pm}e^{-t}\left(\cosh(t\Delta)\pm
(c_+-c_-){\sinh(t\Delta)\over \Delta}\right),
\end{equation}
where a shorthand notation, $\Delta=\sqrt{1-4c_+c_-(1-w^2)}$, has been
used.  In principle, one then finds $P_n^+(t)$ and $P_n^-(t)$ by
expanding the generating functions. This leads to rather cumbersome
results.  However, the most interesting scaling results correspond to
the limit $t\to \infty,\,\, 1-w\to +0$ with $(1-w)t$ kept finite.  In
this scaling limit, $1-\Delta\to 4c_+c_-(1-w)t$. Substituting this into
Eq.~(\ref{gen}) we find $F_{\rm even}^{\pm}\simeq
c_+^2\exp(-4c_+c_-(1-w)t)$.  Then we find $F_{\rm odd}^{\pm}$, note
that in the scaling limit the generating functions become the Laplace
transforms of $P_n^+(t)$ and $P_n^-(t)$, and perform the inverse
transformation.  Finally, we arrive at the following scaling results
\begin{eqnarray}
\label{final}
{P_{2n}^+\over c_+^2}\simeq {P_{2n}^-\over c_-^2}&\simeq&
{P_{2n+1}^+\over c_+c_-}\simeq {P_{2n+1}^+\over c_+c_-}\simeq \nonumber\\
&&{1\over\sqrt{2\pi c_+c_-t}}\,\exp\left[-{(n-2c_+c_-t)^2\over
2c_+c_-t}\right].
\end{eqnarray}
In particular, we see that for
$c_+\neq c_-$ the distribution function for the even number of
changes, $P_{2n}=P_{2n}^++P_{2n}^-$, is larger than the distribution
function for the odd number of changes, $P_{2n+1}=
P_{2n+1}^++P_{2n+1}^-$.  Eq.~(\ref{final}) suggests that it is
possible to avoid these ``even-odd oscillations'', by making a
transformation to a modified opinion change distribution $\tilde
P_n\equiv P_n+P_{n+1}$.  We also note that the scaling distribution in
the right hand side of Eq.~(\ref{final}) is identical with the
infinite-dimension scaling function, previously obtained for the
symmetric case.

\subsection{Exact Results}

Although the above results were obtained using mean-field
considerations, similar behavior characterizes the exact solution.  By
generalizing the solution of Eq.~(\ref{auto}), the autocorrelation
function is found
\begin{equation}
\label{autoex}
A(t)=\sum(-1)^nP_n=(c_+-c_-)^2+4c_+c_-[I_0(t)e^{-t}]^d.
\end{equation}
The limiting value of the autocorrelation function, $(c_+-c_-)^2$, is
identical with the mean-field theory Eq.~(\ref{automf}). Again the
conclusion remains the same, at the late stages of the process a
single voter opinion cannot be used to determine its initial
opinion. Similar to the symmetric case, the
autocorrelation function decays algebraically rather than
exponentially with time.
Since $P_{\rm even}=(1+A(t))/2\geq P_{\rm odd}=(1-A(t))/2$,
we also learn that a voter is more likely to have its initial
opinion.

Mean-field theory suggests that the fraction of persistent voters
decays faster for the minority. It is interesting to investigate the
same for the one-dimensional situation.  It is instructive to start
with the special case of $c_+=1/3$ and $c_-=2/3$.  Let us formally
split the $-$ opinion into two equivalent sub-opinions.  Hence, there
are three equiprobable opinions, one $+$ opinion and two $-$
sub-opinions.  We now identify this system as the zero-temperature
three states Potts model, or as a voter model with three opinions. The
dynamics is unchanged, a voter chooses a nearest neighbor randomly,
and assumes its opinion. Eventually, we will not distinguish between
the $-$ sub-opinions.  For the kinetic $q$-state Potts model with
$T=0$, the fraction of persistent spins decays according to
$P_0(t)\sim t^{-\beta(q)}$, with
$\beta(q)=2\pi^{-2}\left[\cos^{-1}(\sqrt{2}q^{-1}-1/\sqrt{2})\right]^2-1/8$,
see \cite{der}.  Indeed, for the symmetric voter model, $q=2$
and $\beta(2)=3/8$. The concentration of persistent minority species,
$P_0^{+}(t)$, is equal to the fraction of persistent spins in the
$q$-state Potts model with $q=3$. Using the notation $P_0^{\pm}(t)\sim
t^{-\beta_{\pm}}$, one has $\beta_+=\beta(3)\cong 0.5379$.  Of course,
$\beta_-\neq\beta_+$, since changes between $-$ sub-opinions should not
be counted. The exponent $\beta_-$ can be found by allowing a
non-integer number of opinions, $q=1/c_-=3/2$. This formula is found
by an analytical continuation to arbitrary $q$ of the relation $c=1/q$,
which clearly holds in the equal-concentration case with an integer
$q$. Therefore, $q_{\pm}=1/c_{\pm}$.  For the above example,
$c_-=2/3$, $q_-=3/2$ implies $\beta_-\cong0.2349$.  In general, the
concentration of persistent voters decays algebraically
\begin{equation}
\label{p0beta}
P_0^{\pm}\sim t^{-\beta_\pm}\quad{\rm with}\quad \beta_\pm=
{2\over\pi^2}
\left[\cos^{-1}(\sqrt{2}c_\pm-1/\sqrt{2})\right]^2-{1\over 8}.
\end{equation}

Following Eq.~(\ref{1dscaling}), $P_n(t)$ can be written in terms of a
simple scaling function in one dimension.  The $z\to 0$ behavior reflects the
anomalously large number of persistent voters found in the system at
long times. On the other hand, Eq.~(\ref{p0beta}) implies a difference
in nature of the scaling functions for sites of initial $+$ and $-$ opinion,
$P_n^{\pm}(t)=\Phi_1^{\pm}(n/\sqrt{t})/(c_{\pm}\sqrt{t})$. In the
limit of large $z=n/\sqrt{t}$, the tail is dominated by Gaussian
fluctuations, while in the limit $z\to 0$, the anomalous
decay of Eq.~(\ref{p0beta}) determines the behavior. Combining these
two limits, we have
\begin{equation}
\label{1dlimitsuneq}
\Phi_1^{\pm}(z)\sim \cases
{z^{2\beta({c_\pm})-1}&$z\ll 1;$\cr
\exp(-{\rm const}\times z^2)&$z\gg1$.\cr}
\end{equation}
In the limit of a vanishing minority opinion concentration, $c_+\to
0$, one has $\beta_+\to 1$, and $\beta_-\cong 2c_+/\pi\to 0$.

Both the mean-field results and our numerical simulations, to be
described in the following, suggest that distribution of even number
of changes dominates over its odd counterpart. We expect that the
above suggested scaling form holds for the even distribution, or
equivalently, for the modified distribution $P_n+P_{n+1}$.
To summarize, the exact form of the fraction of
persistent voters combined with scaling considerations suggest that
different scaling functions correspond to the minority and the
majority opinions.

\subsection{Infinitesimal concentrations}

For better understanding of the asymmetric case, it is useful to
consider the case of an infinitesimal concentration of one opinion,
$c_+\to 0$. We naturally restrict ourself to the situation where a
single $+$ voter is placed in a sea of $-$ opinion.  Identifying an
interface between $+$ and $-$ domains with a random walker, an
equivalence to two annihilating random walkers who are nearest
neighbors at $t=0$, is established.  The distribution $P_n(t)$ is thus
equal to the fraction of sites visited $n$ times by the two walkers.  We
further simplify the problem by considering the fraction of sites
visited by a single random walk with a trap as one of its nearest
neighbors.  Although the two problems are not identical, we expect
that the results are similar in nature and differ only by numerical
prefactors. The reason is that the distance between the two random
walks itself performs a random walk with in the vicinity of a trap.

In the limit of a vanishing opinion concentration, $c_+\to 0$, the
opinion change density $P_n(t)$ is equal to zero. However, if we
divide $P_n^-(t)$ by the density of the interfaces, $c_+c_-$, and then
go to the limit $c_+\to 0$, we obtain a nontrivial distribution,
$\lim_{c_+\to 0}P_n^-(t)/c_+c_-$. This distribution gives the total
number of links crossed $n$ times by the walker; we will denote it by
$P_n(t)$.

As said previously, for the symmetric initial conditions,
$c_+=c_-=1/2$, the scaling behavior of the form $P_n(t)=t^{-1/2}
\Phi_1\left(n/\sqrt{t}\right)$ is expected. However, for
asymmetric initial conditions, two different scaling forms, even and
odd, should appear.  In the present extreme case, we expect
$P_{2n}(t)=t^{-1/2}\,\Phi_{\rm even}\left(n/\sqrt{t}\right)$ and
$P_{2n+1}(t)=t^{-1/2}\,\Phi_{\rm odd}\left(n/\sqrt{t}\right)$.
We learn from Eq.(\ref{1dlimitsuneq}) that $\Phi_{\rm even}\equiv\Phi_1^-\sim
z^{-1}$ near the origin. Hence, the distribution function approaches a
time independent form: $\lim_{t\to\infty}P_n(t)\sim n^{-1}$.

These results can be confirmed by considering the analogy to a single random
walk near a trap.  As the walker will ultimately come to the origin
with probability one, every link $(k-1,k),\, k\geq 2$ will be crossed
an even number of time and so the ultimate distribution
$P_{2n+1}(\infty)=0$ for $n\geq 1$ (and $P_1(\infty)=1$ since the link
$(0,1)$ is crossed with ultimate probability one by the walker).  So,
in the extreme case we are considering, the even-odd oscillations are
obvious and pronounced: The asymptotic even values are positive while
the odd values are zero.

In order to compute $P_n(\infty)$ for $n$ even, we consider the link
$(k-1,k)$. The probability that the walker starting at $x=1$ reaches
for the first time $x=k$, thus crossing the link $(k-1,k)$, is given
by $p(k)=1/k$ \cite{rubin}.  Then the ultimate probability that the
walker will go from site $x=k$ to site $x=k-1$, crossing the link
$(k-1,k)$ a second time, is one.  The probability that the walker
starting at $x=k-1$ will arrive at $x=0$ before crossing the link
$(k-1,k)$ again, is given by $1/k$.  Therefore, $k^{-2}$ is the
contribution of the link $(k-1,k)$ into $P_2(\infty)$, the average
number of links crossed twice by the walker.  Thus, we have
\begin{equation}
      \label{p2}
P_2(\infty)=\sum_{k=2}^\infty {1\over k^2}= \zeta(2)-1
={\pi^2\over 6}-1.
\end{equation}
After having crossed the link $(k-1,k)$ twice, the walker could
cross this link again before reaching the adsorbing barrier at
$x=0$.  Any such crossing from the left happens with probability
$1-1/k$, while the next crossing from the right happens with
probability one.  Thus, we arrive at the remarkably simple formula
expressing $P_n(\infty)$ through the zeta function
\begin{eqnarray}
      \label{pn}
P_{2n+2}(\infty)&=&\sum_{k=2}^\infty
\left(1-{1\over k}\right)^n{1\over k^2}\nonumber\\
&=&\sum_{m=0}^n\left(\begin{array}{c}
                            n\\m
                  \end{array}\right)
(-1)^m \left(\zeta(m+2)-1\right)
\end{eqnarray}
For large $n$, the sum can be approximated by the integral
\begin{equation}
      \label{pnap}
P_{2n}(\infty)\simeq\int_{0}^{1/2}(1-\xi)^{n-1}\, d\xi=
{1-2^{-n}\over n}\simeq {1\over n}
\end{equation}
which confirms the above prediction.

To determine the scaling functions $\Phi_{\rm even}(z)$ and $\Phi_{\rm
odd}(z)$, it proves useful to consider $P_n(x,t)$, the probability
that the walker passes $n$ times through $x$ during the time interval
$(0,t)$.  The $P_n(t)$ is then given by
\begin{equation}
  \label{cont}
P_n(t)=\sum_{x=2}^\infty\, P_n(x,t) \simeq \int_2^\infty dx\, P_n(x,t).
\end{equation}
In this equation and in the following we will treat $x$ as a
continuous variable; in the long-time limit, this should be
asymptotically correct.

We then write for $P_n(x,t)$:
\begin{eqnarray}
  \label{even}
P_{2n}(x,t)&=&\int_0^t dt_1 p_1(x,t_1)\int_0^{t-t_1}
dt_2 p_2(t_2)\int_0^{t-t_1-t_2} dt_3 p_3(x,t_3) \nonumber\\
&\ldots &\int_0^{t-\sum_{i\leq 2n-1} t_i} dt_{2n} p_2(t_{2n})\,
p_4\left(x,t-\sum_{i=1}^{2n} t_i\right)
\end{eqnarray}
and
\begin{eqnarray}
  \label{odd}
P_{2n+1}(x,t)&=&\int_0^t dt_1 p_1(x,t_1)\int_0^{t-t_1}
dt_2 p_2(t_2)\int_0^{t-t_1-t_2} dt_3 p_3(x,t_3) \nonumber\\
&\ldots &\int_0^{t-\sum_{i\leq 2n} t_i} dt_{2n+1}
p_3(x,t_{2n})\,p_5\left(t-\sum_{i=1}^{2n+1} t_i\right).
\end{eqnarray}
We consider a walker starting at $y_0=1$;
$p_1(x,t)$ is the probability that this walker
reaches $y=x$ at time $t$ without going to the origin $y=0$; $p_2(t)$
is the probability that this walker first reaches the origin
at time $t$; $p_3(x,t)$ is the probability that this walker
first passes at the origin at time $t$ without passing through
$y=x$; $p_4(x,t)$ is the probability that this walker
with an absorbing boundary at $y=x$ does not pass through the origin
up to time $t$ and $p_5(t)$ is the probability that this walker does not
reach the origin up to time $t$.  Eq.~(\ref{even}) is
cumbersome in form but simple in nature: The formula for $P_{2n}(x,t)$
is just a finite-time generalization of Eq.~(\ref{pn}), namely it
corresponds to the situation when a walker has performed $n$
oscillations around the link $(x-1,x)$, and at time $t$ a walker, or
his remains, is to the left of $x$.  Eq.~(\ref{odd}) has been
constructed similarly and describes the situation with a walker to the
right of $x$ at time $t$.  The convolution structure of
Eqs.~(\ref{even}) and (\ref{odd}) suggests to apply the Laplace
transform.  Indeed, $\tilde P_n(x,s)=\int_0^\infty dt e^{-st}
P_n(x,t)$, satisfy
\begin{equation}
  \label{evenlap}
 \tilde P_{2n}(x,s)= \tilde p_1(x,s) \left(\tilde p_2(s)\right)^{n}
\left(\tilde p_3(x,s)\right)^{n-1}\tilde p_4(x,s)
\end{equation}
and
\begin{equation}
  \label{oddlap}
 \tilde P_{2n+1}(x,s)= \tilde p_1(x,s) \left(\tilde p_2(s)\right)^{n}
\left(\tilde p_3(x,s)\right)^{n}\tilde p_5(s).
\end{equation}
Fortunately, the probabilities $\tilde p_j$ have been already
computed \cite{feller}:
\begin{eqnarray}
  \label{fs}
\tilde p_1(x,s) & = & {\hbox{sh}}(\sqrt{s})
\over \hbox{sh}(x\sqrt{s})\nonumber\\
\tilde p_2(s) & = & e^{-\sqrt{s}}\nonumber\\
\tilde p_3(x,s) & = & {\hbox{sh}}((x-1)\sqrt{s})\over \hbox{sh}(x\sqrt{s})\\
\tilde p_4(x,s) & = & {1-\tilde p_3(x,s)\over s}\nonumber\\
\tilde p_5(s) & = & {1-\tilde p_2(s)\over s}.\nonumber
\end{eqnarray}
It is in principle possible now to compute various $P_n(t)$.
For example, the contribution to $P_1(t)$ from links with $k\geq 2$ is
\begin{eqnarray}
  \label{p1}
\tilde P_1(s)-\tilde P_1(1,s)&=&{1-e^{-\sqrt{s}}\over s}\hbox{sh}(\sqrt{s})
\int_2^\infty {dx\over \hbox{sh}(x\sqrt{s})}\nonumber\\
&=&{1-e^{-\sqrt{s}}\over s^{3/2}}\hbox{sh}(\sqrt{s})
\ln(\hbox{cth}(\sqrt{s}))\\
&\simeq &{\ln(1/s)\over 2\sqrt{s}}\qquad (s\rightarrow 0),\nonumber\\
\end{eqnarray}
where the contribution from the first link $(0,1)$ is
$P_1(1,t)=1-1/\sqrt{\pi t}$,
which gives the asymptotic value of
\begin{equation}
  \label{asym}
P_1(t)\simeq 1-1/\sqrt{\pi t}+{\ln t\over\sqrt{4\pi t}}
\qquad (t\rightarrow \infty).
\end{equation}

We now turn to determination of the scaling functions. In the
long-time limit, $(t \to \infty)$, corresponding to $(s \to 0)$,
Eq.~(\ref{evenlap}) becomes
\begin{equation}
  \label{evenxs}
 \tilde P_{2n}(x,s)\simeq {e^{-n\sqrt{s}}\over s}\,{1\over x^2}
\left(1-{1\over x}\right)^{n-1},
\end{equation}
which then implies
\begin{equation}
  \label{evens}
 \tilde P_{2n}(s)= \int_2^\infty dx \tilde P_{2n}(x,s)
\simeq {e^{-n\sqrt{s}}\over ns}.
\end{equation}
Performing the inverse Laplace transform \cite{abram}, one finds
\begin{equation}
  \label{escaling}
P_{2n}(t)={1\over n}\,{\rm Erfc}\left({n\over \sqrt{4t}}\right).
\end{equation}
Indeed the anticipated scaling behavior suggested earlier is confirmed
with the scaling function
\begin{equation}
\Phi_{\rm even}(z)=z^{-1}{\rm Erfc}(z/2).
\label{escaling1}
\end{equation}
In particular, the limiting forms are
\begin{equation}
  \label{easymp}
P_{2n}(t)\simeq \cases{
{1\over n}-{1\over \sqrt{\pi t}},  \qquad & $n\ll \sqrt{t}$, \cr
{1\over n^2}\sqrt{4t\over\pi}\,\exp\left(-{n^2\over 4t}\right),
\qquad & $n\gg \sqrt{t}$. \cr}
\end{equation}

For the odd distribution, a similar scaling form is expected:
\begin{equation}
  \label{oddscaling}
P_{2n+1}(t)={1\over \sqrt{t}}\Phi'_{\rm odd}\left({t\over n^2}\right).
\end{equation}
When $(s\to 0)$, we can use
the naive expansion as previously but we should keep the upper limit
finite, $\leq s^{-1/2}$, since the integrand logarithmically diverges
on the upper limit:
\begin{eqnarray}
  \label{odds}
 \tilde P_{2n+1}(s)&\simeq & {e^{-n\sqrt{s}}\over \sqrt{s}}
\int_2^{s^{-1/2}} {dx\over x}\,\left(1-{1\over x}\right)^n \nonumber\\
&\simeq & {e^{-n\sqrt{s}}\over \sqrt{s}}\,E_1(n\sqrt{s}),
\end{eqnarray}
with the exponential integral $E_1(y)=\int_y^{\infty} du
u^{-1}\exp(-u)$.  Making use of Eq.~(\ref{oddscaling}) one gets
another relation for $\tilde P_{2n+1}(s)$,
\begin{eqnarray}
  \label{odds2}
\tilde P_{2n+1}(s)&=&\int_0^\infty dt e^{-st}
{1\over \sqrt{t}}~\Phi'_{\rm odd}\left({t\over n^2}\right)\nonumber\\
&=&n \int_0^\infty {dT\over \sqrt{T}}
e^{-qT}\Phi'_{\rm odd}(T),
\end{eqnarray}
with $q=n^2s$.

Thus we obtain the Laplace transform of the function
$\Phi'_{\rm odd}(T)/\sqrt{T}$,
\begin{equation}
  \label{odds3}
\int_0^\infty {dT\over \sqrt{T}} e^{-qT}\Phi'_{\rm odd}(T)=
{e^{-\sqrt{q}}\over \sqrt{q}}\int_{\sqrt{q}}^\infty {dy\over y}\,e^{-y}.
\end{equation}
Performing the inverse Laplace transform, we get
\begin{equation}
  \label{oddfinal}
{\Phi'_{\rm odd}(T)\over \sqrt{T}}=
\int_0^T {d\tau\over 2\tau}\,{\rm Erfc}\left({1\over \sqrt{4\tau}}\right)\,
{1\over \sqrt{\pi(T-\tau)}}\,
\exp\left(-{1\over 4(T-\tau)}\right).
\end{equation}
Performing asymptotic analysis yields
\begin{equation}
  \label{oasymp}
P_{2n+1}(t)\simeq \cases{
{\ln(t/n^2)\over \sqrt{4\pi t}},  \qquad & $n\ll \sqrt{t}$, \cr
{1\over n^2}\sqrt{2t\over\pi}\,\exp\left(-{n^2\over t}\right),
\qquad & $n\gg \sqrt{t}$. \cr}
\end{equation}
Notice that in the both limiting cases, $P_{2n+1}(t) \ll P_{2n}(t)$.

It proves insightful to compute the moments of even and odd
distributions, $M_{\rm even}^p(t)=\sum_{n\geq 1}(2n)^p P_{2n}(t)$
and $M_{\rm odd}^p(t)=\sum_{n\geq 0}(2n+1)^p
P_{2n+1}(t)$. Asymptotically, it is easy to compute even moments
\begin{equation}
  \label{evenmoment}
M_{\rm even}^p(t)= E_p t^{p/2}, \quad
E_p={4^p\Gamma\left({p+1\over 2}\right)\over p\sqrt{\pi}}.
\end{equation}
Eq.~(\ref{evenmoment}) is valid only for $p>0$ (when $(p\to +0)$, the
prefactor $E_p$ diverges). To determine the most interesting zero
moment, i.e. the total number of links crossed even times $M_{\rm
even}^0(t)=\sum_1^\infty P_{2n}(t)$, we use the the Laplace transform
of Eq.~(\ref{evens}) to obtain $M_{\rm even}^0(s)\simeq
\ln(1-\exp(-\sqrt{s}))/s\simeq \ln (1/s)/2s$ and eventually,
\begin{equation}
  \label{evenzero} M_{\rm even}^0(t)\sim (\gamma+\ln t)/2,
\end{equation}
with $\gamma\cong 0.577215$ the Euler constant.  This result is
consistent with a direct summation of $P_{2n}(\infty)=n^{-1}$ up to
$n=\sqrt{t}$.  For negative $p$, even moments are finite, $M_{\rm
even}^p(\infty)=2^p\zeta(1-p)$.

Odd moments behave similarly, $M_{\rm odd}^p(t)= O_p t^{p/2}$.
A lengthy computation gives the prefactor
\begin{equation}
  \label{oddmoment}
O_p={2^{2p+1}\Gamma\left(1+{p\over 2}\right)\over \pi}
\int_0^1 d\mu\left(1-\mu^2\right)^{p-1\over 2}
\int_0^\mu d\theta{\theta^{p+1}\over 1-\theta^2}.
\end{equation}
Eq.~(\ref{oddmoment}) is valid for all nonnegative $p$,
and in particular the (average) total number of links crossed odd
times approaches a surprising constant
\begin{equation}
  \label{oddzero} M_{\rm even}^0=-{1\over 2\pi}\int_0^1
d\mu\,{\ln(1-\mu^2)\over \sqrt{(1-\mu^2)}}=\ln 2.
\end{equation}
Thus although the odd part of the $P_n$ distribution approaches
zero as $t\to \infty$, the moments remain nontrivial.

\subsection{Simulation Results}

To test the above predictions we performed numerical simulations of
the voter model with different initial concentrations, in
one dimension. The rich behavior predicted by the mean-field and the
exact results was confirmed by the simulation results.
We studied the fraction of persistent voters for the case
$c_+=1/3$, and we found the decay exponents $\beta_+=0.54$, and
$\beta_-=0.23$ for the minority and the majority opinion,
respectively. These values are in excellent
agreement with Eq.~(\ref{p0beta}).

We also confirmed that each of the four functions $P_{2n}^{\pm}(t)$
and $P_{2n+1}^{\pm}$ can be rewritten in a  scaling  with the
scaling variable $n/\sqrt{t}$.  The dominance of the even part of the
distribution $P_{2n}>P_{2n+1}$, is nicely demonstrated by Fig.~(4)
(one realization of a system of $10^6$ sites) and the asymptotics of
the even scaling function Eq.~(\ref{1dlimitsuneq}) are verified.

We performed also simulations for the extreme case $c_+\to 0$,
where one site with initial opinion $+$ is in a sea of $-$ opinions.
As shown above, this problem is equivalent to the average number of times
a link is crossed by two annihilating random walkers.
We show on Fig.~(4) the even and odd scaling functions for $10^8$
realizations of this system.
The asymptotic results Eqs.(\ref{easymp}) found in the simplified
problem of one random walker in the presence of an absorbing
boundary conditions are verified up to numerical prefactors.
In particular, the even scaling functions of Fig.~(4) is found to behave
asymptotically ($z\to 0$) as $\Phi_{\rm even}(z)\simeq 5/(4z)$ to be compared
with $\Phi_{\rm even}(z)\simeq 1/z$ of Eq.~(\ref{escaling1}).

\section{Summary}

We have investigated the voter model, one of the simplest models of
non-equilibrium statistical mechanics with {\it non-conserved}
dynamics.  We have introduced the set of quantities $P_n(t)$, defined as
the fraction of voters who changed their opinion $n$ times up
to time $t$.  The distribution $P_n(t)$ was shown to exhibit a
scaling behavior that strongly depends on the dimension of the system
and on the opinion concentrations.  For $d>2$, the system does not
coarsen, and the distribution is Poissonian.  In one-dimension, We have
solved for $P_n(t)$ in the extreme case when the minority opinion is
infinitesimal. The case when the minority phase occupies a negligible
volume has been studied in the classical work \cite{lif} for the {\it
conserved} dynamics and has proven very important in the development
of the theory of phase ordering kinetics \cite{bray}. It would be very
interesting to generalize the extreme-case solution to arbitrary $d$.

The quantity $P_n(t)$ reflects the history of the coarsening process.
Knowledge of this distribution enables insight into interesting
quantities such as the fraction of consistent or ``frozen'' sites, the
fraction of sites with their original opinion, and the average number
of changes in a site.  This study suggests that $P_n(t)$ is a tool for
investigations of coarsening processes in more realistic models.  It
possible that a Poissonian  $P_n(t)$ generally describes systems that do
not coarsen, while asymmetric distributions which are pronounced
near the origin  correspond to coarsening systems.

\section*{Acknowledgements}

It is a pleasure to thank S. Redner for fruitful discussions.
E.~B.~ was supported in part by NSF under Award Number 92-08527 and by the
MRSEC Program of the National Science Foundation under Award Number
DMR-9400379. L.~F. was supported by the Swiss National Science Foundation.

\section*{Figure Captions}

\noindent Fig. 1 Scaling for the symmetric case in one-dimension. The
       quantity $t^{1/2}P_n(t)$ is plotted versus $n/t^{1/2}$ for
       different times, $t=10^3,10^4,10^5$.

\noindent Fig. 2 The distribution function $P_n(t)$ versus $n$ at time $t=2000$
       in 2D.

\noindent Fig. 3 The fraction of persistent voters in 2D, $P_0(t)$ versus
       $\ln^2 t$.  An average over 300 samples of linear size $L=10^3$
       for $c_+=c_-=1/2$ (solid line) and over 50 samples of linear
       size $L=10^3$ for $c_+=1-c_-=1/4$ (dashed line).

\noindent Fig. 4 The even and odd distribution functions for different
       $c_+$ in 1D.  $t^{1/2}P_n(t)/(c_+c_-)$ is plotted versus
       $n/t^{1/2}$.  Different scaling functions correspond to the
       even (upper curves) and the odd (lower curves) parts of the
       distribution. The solid lines correspond to the case $c_+=1/4$
       for one sample of linear size $L=10^6$.  The dashed lines
       correspond to the

\begin{figure}
\centerline{\epsfbox{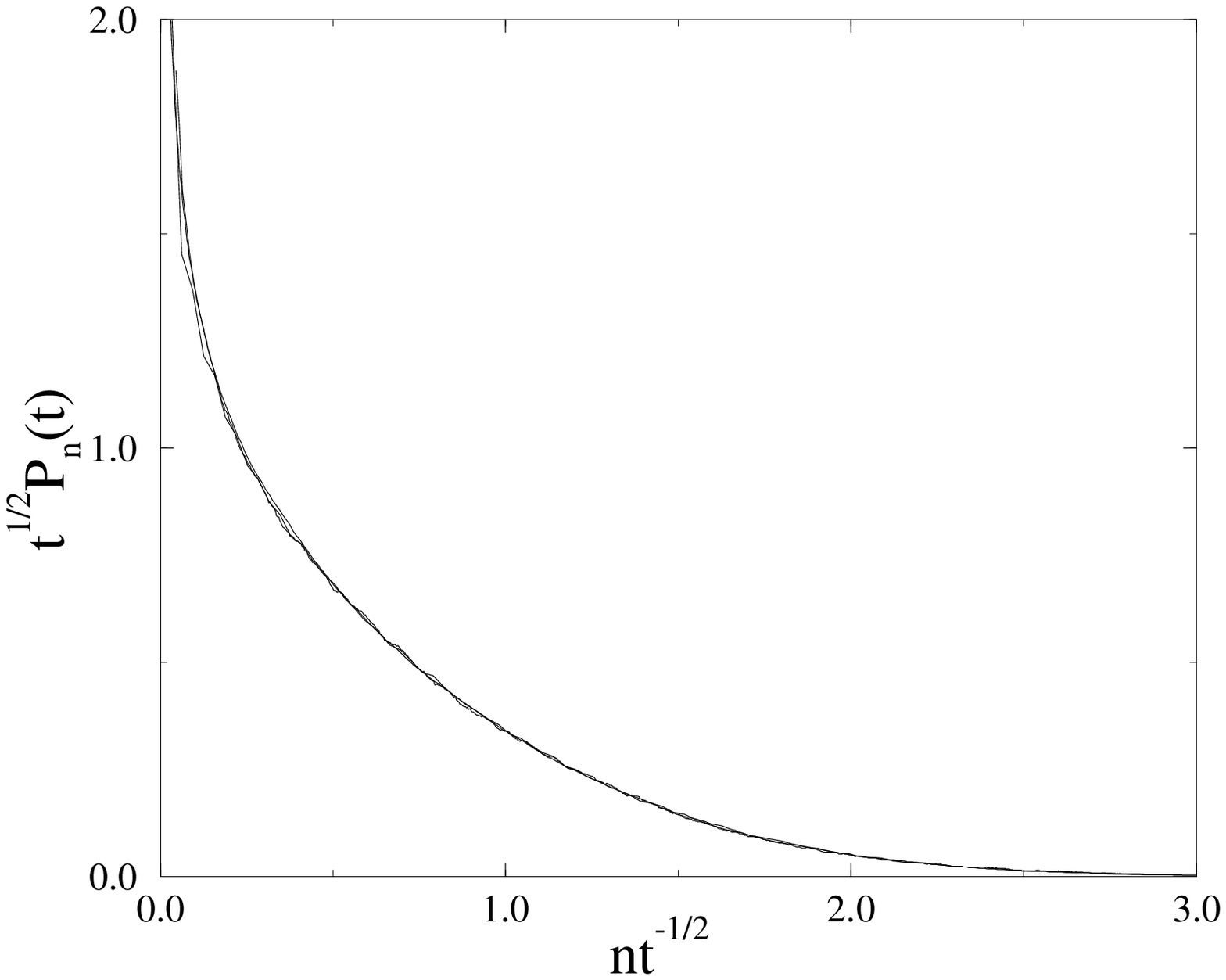}}
Figure 1
\end{figure}

\begin{figure}
\centerline{\epsfbox{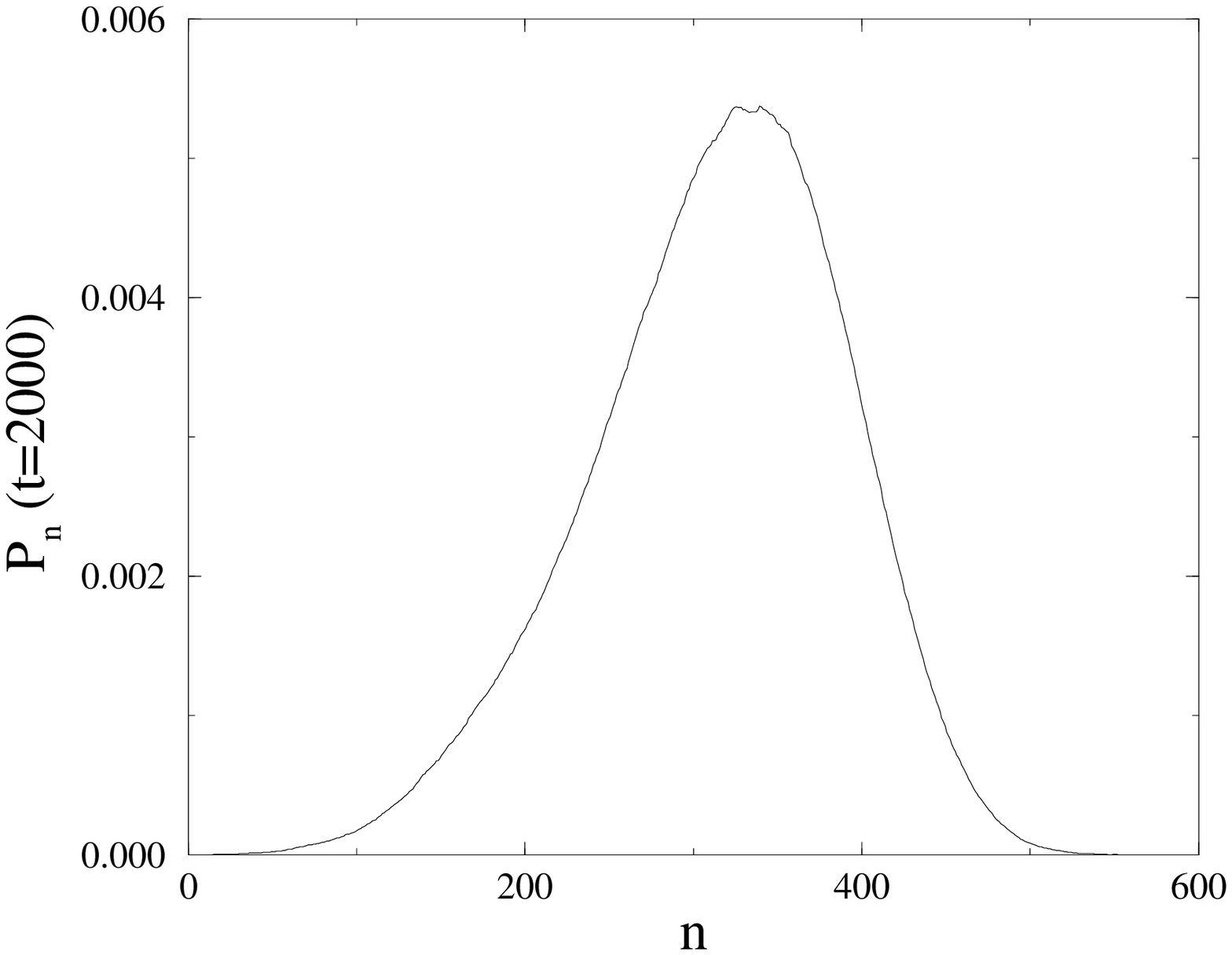}}
Figure 2
\end{figure}

\begin{figure}
\centerline{\epsfbox{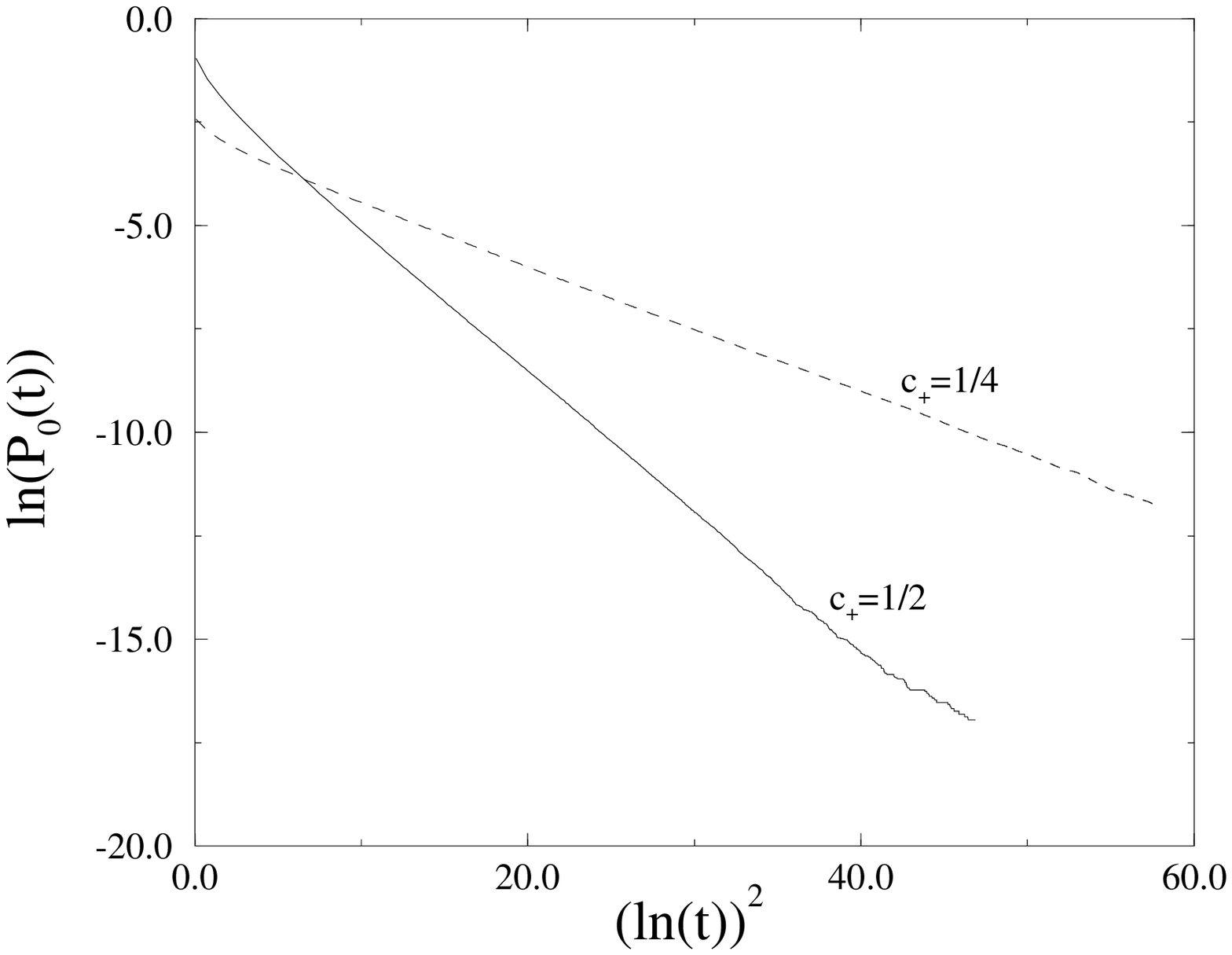}}
Figure 3
\end{figure}

\begin{figure}
\centerline{\epsfbox{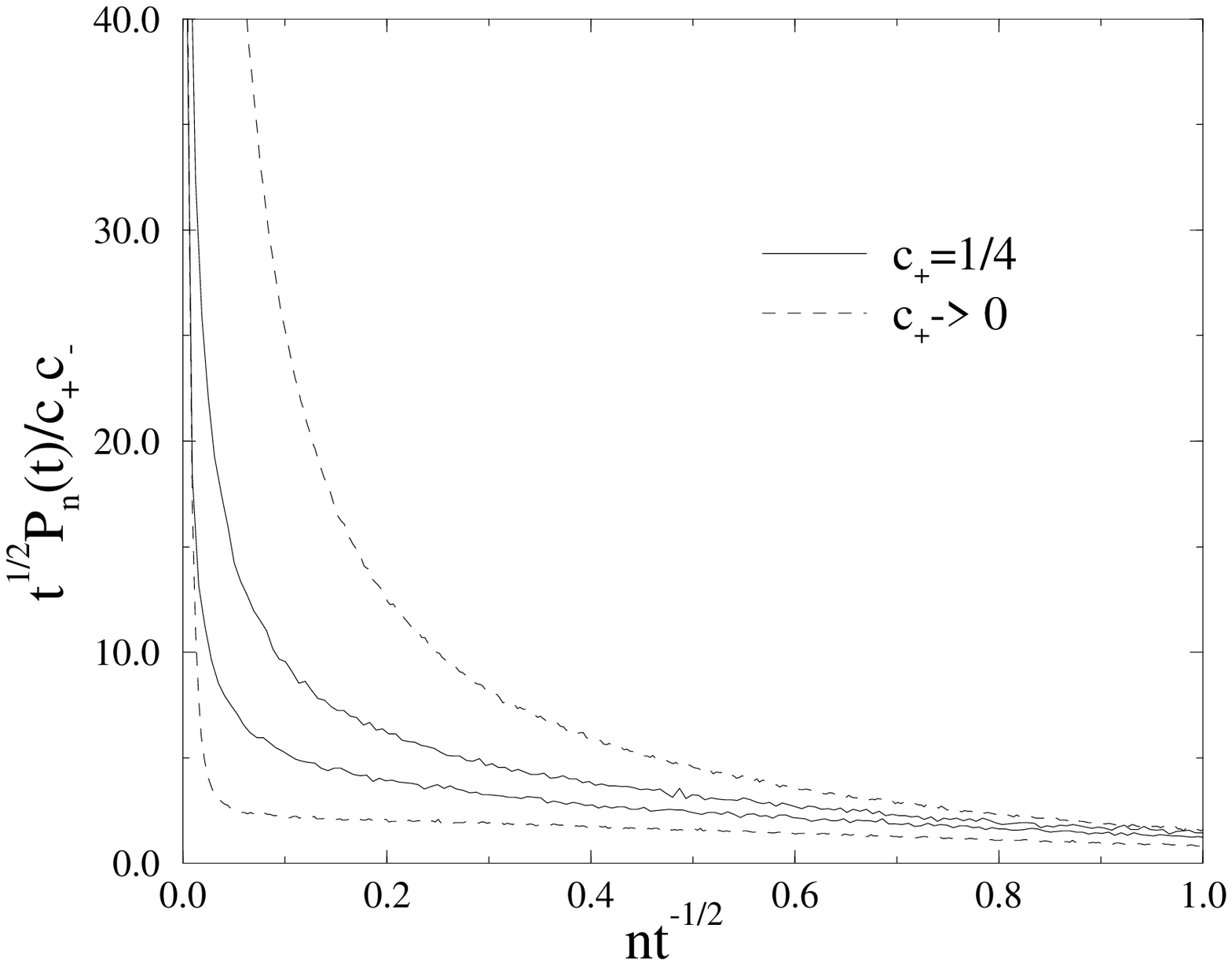}}
Figure 4
\end{figure}

\end{document}